# Evidence of both surface and bulk Dirac bands in ZrSiS and the unconventional magnetoresistance

Xuefeng Wang,[1,*] Xingchen Pan,[2] Ming Gao,[1] Jihai Yu,[2] Juan Jiang,[3] Junran Zhang,[1] Huakun Zuo,[4] Minhao Zhang,[1] Zhongxia Wei,[2] Wei Niu,[1] Zhengcai Xia,[4] Xiangang Wan,[2] Yulin Chen,[3] Fengqi Song,[2,†] Yongbing Xu,[1] Baigeng Wang,[2] Guanghou Wang,[2] and Rong Zhang[1,‡]

[1]National Laboratory of Solid State Microstructures, Collaborative Innovation Center of Advanced Microstructures, and School of Electronic Science and Engineering, Nanjing University, Nanjing 210093, China

[2]National Laboratory of Solid State Microstructures, Collaborative Innovation Center of Advanced Microstructures, and College of Physics, Nanjing University, Nanjing 210093, China

[3]School of Physical Science and Technology, ShanghaiTech University, and CAS-Shanghai Science Research Center, Shanghai 200031, China

[4]Wuhan National High Magnetic Field Center, Huazhong University of Science and Technology, Wuhan 430074, China

Corresponding authors.

[*]xfwang@nju.edu.cn, [†]songfengqi@nju.edu.cn, [‡]rzhang@nju.edu.cn.

**Abstract**

The unconventional magnetoresistance of ZrSiS single crystals is found unsaturated till the magnetic field of 53 T with the butterfly shaped angular dependence. Intense Shubnikov-de Haas oscillations resolve a bulk Dirac cone with a nontrivial Berry phase, where the Fermi surface area is $1.80 \times 10^{-3}$ Å$^{-2}$ and reaches the quantum limit before 20 T. Angle resolved photoemission spectroscopy measurement reveals an electronic state with the two-dimensional nature around the X point of Brillouin zone boundary. By integrating the density functional theory calculations, ZrSiS is suggested to be a Dirac material with both surface and bulk Dirac bands.

Recently, mining the topological anomaly in the electronic structures and the transport signature has been a central topic in the community of quantum materials [1-8]. Some new concepts are emerging including topological insulators, topological Dirac semimetals, and Weyl semimetals, all of which have demonstrated novel transport properties with the potential device applications [1,9-13]. Most of topological insulators and Weyl/Dirac semimetals are featured with a linear/quasi-linear dispersion in their band structures, which drives modulated electronic accumulation in the lowest Landau level and paves the dominance of linear magnetoresistance (MR) in high magnetic fields [7,8,14-17]. The large Berry curvature exists in the bands of TaAs and NbP, which casts a large room-temperature MR ratio of up to 100,000 [17,18]. It can also be related to the unexpectedly high room-temperature mobility of these materials. The opposite Weyl charges can be found in the Weyl/Dirac semimetals, which can simulate the chiral anomaly of quasiparticle transport proposed in high energy physics. It leads to the intense negative MR when the field is aligned along the direction of the current, which vanishes totally when the field is tilted by a very small angle [18-21]. Perfectly compensated electron/hole pockets have been found on the Fermi level of $WTe_2$, leading to the classic MR resonance and an unsaturated MR till the large field of 53 T [2]. In $WTe_2$, the intriguing linear MR appears when the field is aligned both in the perpendicular and parallel directions [22-24]. When the large MR is suppressed by a pressure, superconductivity arises due to the possible Fermi level nesting [25,26]. The similar physics generating a large quadratic MR of up to over 10,000 times has been observed at 9 T in TaAs [27]. Recently, the type-II Weyl semimetal has been studied [28] and confirmed by visiting the empty states using the laser pumping [29].

Intense interest paves the way to further investigate the aforementioned topic. ZrSiS, with the iron-pnictide superconductor LiFeAs structure, has been predicted to be a two-dimensional (2D) oxide topological insulator [30] and host several bulk Dirac cones [31]. In this work, combining the studies of high field magnetotransport till 53 T, angle-resolved photoemission spectroscopy (ARPES) and density functional theory (DFT) calculations, we provide the unambiguous evidence of the ZrSiS as a new type of quantum material with the Dirac cones from both the bulk and surface electronic states. The unconventional MR keeps unsaturated till the intense field of 53 T and depicts the interesting butterfly shaped angular dependence.

The single crystals were grown by a chemical vapor transport approach, as described elsewhere [31]. More details are shown in the Supplemental Material [32]. To study the magnetotransport properties, hall-bar and four-terminal devices were fabricated firstly. Then the low-field transport properties were measured by a Physical Property Measurement System (PPMS-9T, Quantum Design), and high-field measurements were performed in a pulsed magnetic field (up to 53 T) at Wuhan National High Magnetic Field Center, China. Due to the substantial aging of the samples, different samples were used in different turns of measurements. ARPES measurements were performed at Beamline 10.0.1 of the Advanced Light Source (ALS) at the Lawrence Berkeley National Laboratory (see the Supplemental Material [32] for details). Density functional theory (DFT) band calculations were performed using WIEN2K code (see the Supplemental Material [32] for details).

Figure 1 shows the transport measurements of the single crystal. The positive temperature-dependent resistance curve indicates the metallic behavior. Unlike other metals, the nonmagnetic material exhibits a strong MR response, as shown in Fig. 1(a), where the resistance increases with the increasing magnetic field. At the temperature of 2 K and the magnetic field of 9 T, the MR ratio reaches nearly 10 000% (Fig. 1(a)). The Hall resistance is measured in a 6-electrode configuration, illustrating a linear trend at high temperatures and a bending feature at temperatures lower than 50 K. This indicates a dominant transport electronic state at high temperatures and more electronic states participating in the transport at lower temperatures. Using a pulsed high field, the MR is found to keep increasing with the magnetic field and shows no saturation when an intense field of 53 T is applied. The MR ratio increases when the temperature decreases and can reach 170 000% at 53 T at 2 K [Fig. 1(a)]. The MR unsaturation is shown more explicitly in the log-log frame, as shown in Fig. 1(c), where no compromise is seen in the MR increase from very low to the extreme field of 53 T. Interestingly, when we measure the MR at a series of configurations with different angles between the field and crystal cleavage plane, it is found that the maximum MR ratio of nearly 200 000% is reached at around 45º [Fig. 1(b)]. Fig. 1(c) also extracted various power factors from 1.44 to 1.8 at different angles. This indicates a strong anisotropy of the MR response. Fig. 1(d) shows the detailed angle-dependent measurement at 5 and 9 T, where the MR ratio of over 10,000% can be achieved at 43 º at 9 T and 2 K. The angle dependent MR shows an interesting butterfly pattern.

The Dirac mode in the bulk ZrSiS is demonstrated by the Shubnikov-de Haas (SdH)

oscillations [33]. Such SdH oscillations become periodic in the frame of $1/B$ and may extract the detailed transport parameters of the electronic states. It may consider the geometric phase of the electronic rotation and forms an effective method to identify the topological-nontrivial transport. As seen in Figs. 2(a) and (c), some oscillations appear and become more apparent at low temperatures. While we analyze the residual signals in the frame of $1/B$ after subtracting the smooth background between 0-9 T [Fig. 2(a)], periodic SdH oscillations appear and support the formation of the Landau levels. There are two sets of SdH oscillations, which can be identified with the frequency of 18.9 T (low-frequency mode) and 246.3 T (high-frequency mode) after the fast Fourier transformation (FFT) [Fig. 2(b)]. Standard SdH analysis gives two electronic states with the Fermi surface areas of $23.5 \times 10^{-3}$ and $1.8 \times 10^{-3}$ Å$^{-2}$, respectively. Cyclotron (effective) mass of the electronic states are 0.12 and 0.16 $m_e$, respectively, by using the Lifshitz-Kosevich formula [Fig. 2(d)]. Both modes exhibit the SdH mobility of over 5,000 cm$^2$/Vs based on the calculation of the quantum transport time of the carriers. The mobility can be much larger since the recent consideration believes the relaxation time can be at least twice of the present values for Dirac materials [7,15]. We note that one SdH mode approaches the quantum limit under high fields, which poses substantial difficulties to the SdH analysis of the pulsed field data. We therefore use the data between 0-9 T and obtains the transport electronic states. Some other parameters are summarized in Supplemental Material [32].

The Landau levels are formed by the quantization of the rotating electrons, which collect the geometric phase and lead to a shift of SdH as indicated by the formula [34],

$$\Delta G = A_0 \frac{\frac{2\pi^2 m^* k_B T}{\hbar e B}}{\sinh\left(\frac{2\pi^2 m^* k_B T}{\hbar e B}\right)} \exp\left(\frac{\pi}{\mu B}\right) \cos\left[2\pi\left(\frac{B_F}{B} - \frac{1}{2} + \beta\right)\right]$$

where $G$ is the conductivity, $m^*$ is the effective mass, $B_F$ is the SdH period and $\beta$ refers to the Berry phase. The nontrivial Berry phase has been a compelling evidence of the Dirac transport, i.e. transport of the electronic states with a Dirac node under the topological protection. In Figs. 2(c) and (d), Landau fan diagrams are constructed by taking the maximum and minimum of the oscillation amplitude $\Delta R_{xx}$ as the half integer and integer, respectively. The $\beta$ factors are 0.04 and 0.5 for high-frequency and low-frequency SdH modes, respectively. Each of them corresponds to an electronic state on the Fermi surface. The Berry phase of 0.5 indicates the nontrivial transport and evidences the Dirac cones. The spin degeneracy of the electronic states may be lifted by the intense fields, which can be evident by the SdH oscillations as seen in the recent study of Dirac materials [35]. Fig. 2(e) shows the MR curve at high fields, where the SdH patterns are obviously split. As transformed to the frame of 1/$B$, one can find most of the SdH waves are split. The split features appear at low temperatures (below 10 K) and high fields (over 20 T), which can be interpreted based on the field-induced gap of spin degeneracy and its competition with the temperature smearing. This is the result of the strong interaction between the electronic state and the magnetic field.

ARPES is a powerful tool to investigate the electronic states of solids. The band structure of ZrSiS obtained from ARPES is shown in Fig. 3. The Fermi surface consists of an elliptical electron pocket around the BZ boundary X and lenses-like hole pockets along ΓM direction, as shown in Fig. 3(a). Interestingly, the electronic state at the X point is absent from bulk calculations. To investigate the nature of this pocket, we conduct the photon energy dependent experiment on both the cut along the ΓX direction and the XM direction in Figs. 3(b) and (c), respectively. The photon energies vary from 30 to 70 eV, and covers more than one BZ along the $k_z$ direction, which is sufficient to illustrate the periodicity along $k_z$. As can be seen in Figs. 3(b) and (c), the band dispersions remain unchanged with different photon energies, which is better illustrated with their corresponding photon energy-$k_{//}$ intensity map shown in Figs. 3(d) and (e). This observation consolidates the 2D nature of the electronic state around X.

The DFT calculations are carried out using the WIEN2K code [36]. In the electronic structures of the bulk solid [Fig. 4(b)], Dirac cones can be found near the Fermi surface, consistent with the previous calculations [30,31]. There are also other Dirac-cone-like features at X and R point below the Fermi level. The Dirac cones along the symmetry line are of small Fermi surfaces, which contribute to the SdH oscillations with a small Fermi surface and a nontrivial Berry phase. The MR anisotropy can be attributed to the anisotropy of the electronic structures. The Fermi velocity determines the quality of carrier transport and is related to the MR ratio. We have calculated the Fermi velocity while we tilt the observation plane from the [001] direction as shown in the inset of Fig. 1(d). As shown in Fig. 4(d), we find that the calculated shows a systematic evolution, where the minimum velocity is near 42º. This helps to understand the butterfly shaped MR anisotropy.

The APRES study indicates the onset of 2D surface states, which could be originated from the quasi-free-standing monolayer on the surface of ZrSiS crystal, as proved by our DFT calculations. This is reasonable since the surface state can be better probed by the ARPES, where its electronic transmission depth is well within several nanometers although its transport is often too weak to be resolved in the bulk-metallic ZrSiS crystal. Furthermore, the lattice spacing between the top layer and adjacent layer is long believed to expand in the surface studies [37], which may lead to the quasi-free-standing monolayer on the ZrSiS surface. This is confirmed by our calculations. The typical ARPES pattern consists of a diamond shaped feature and an elliptical pocket around the X point (Fig. 3(a)). When we compare the calculated Fermi surface of the monolayer and the bulk solid, we find no closed feature near the X point on the Fermi surface of the bulk solid, as shown in the top panel of Fig. 4(c), while it is marked in the monolayer [the bottom panel of Fig. 4(c)]. Furthermore, the band dispersion of this electronic state is also reproduced by the calculation [X in Fig. 4(e)] [31,38]. We further study the electronic structure of the monolayer [Fig. 4(e)], where a $C_{2v}$ symmetry-protected Dirac cone is seen between M and Γ points. The calculation also supports the topological-nontrivial nature of the monolayer [30]. Such topological-nontrivial and Dirac-type bands exist on the surface of ZrSiS crystal, supporting its potential application as 2D weak topological insulators [30,31,38].

In conclusion, ZrSiS hosts the Dirac cones from both the bulk and surface states. The surface Dirac state is a 2D electronic state contributed by the top atomic layers. The bulk Dirac state is verified by the nontrivial Berry phase extracted from the SdH oscillations. The Fermi surface area is $1.80 \times 10^{-3}$ Å$^{-2}$ and reaches the quantum limit before 20 T. The ZrSiS exhibits an unsaturated MR till the intense field of 53 T. The anisotropic Fermi velocity leads to the butterfly shaped angular dependent MR. Our work contributes a new type of Dirac semimetal to the emergent family of quantum materials.

X. F. W. and F. Q. S. gratefully acknowledge the financial support from the National Key Projects for Basic Research of China (Grant No. 2013CB922103 and 2014CB921100), the National Natural Science Foundation of China (Grant Nos. 11274003, 91421109, 11134005, 11522432, 11525417 and 11574288), the Natural Science Foundation of Jiangsu Province (Grant No. BK20130054), the PAPD project, and the Fundamental Research Funds for the Central Universities. Y. L. C acknowledges the support of the EPSRC Platform Grant (Grant No. EP/M020517/1).

X. F. W., X. C. P., M. G., and J. H. Y. contributed equally to this work.

*Note added*. —During the preparation of this manuscript, we became aware of the preprints with the data of large MR in ZrSiS [39,40]. The butterfly MR is seen in arXiv:1603.09318 [40], which also remarks the 2D nature of ZrSiS.

**Figure captions**

**Figure 1. Large, unsaturated and butterfly shaped anisotropic MR of the ZrSiS single crystals.** (a) The MR curves at different temperatures in the field range of 0-53 T. (b) The MR curves using different measurement angles in the field range of 0-53 T at 4.2 K. (c) The MR data shown in the log-log frame, where the MR power can be seen by fitting its slope. The dashed lines show the linear and parabolic increase for eye guiding. (d) Angular dependence of the MR ratio, forming a butterfly shape.

**Figure 2. Analyzing the SdH oscillations for a nontrivial Berry phase.** (a) The relative MR oscillations plotted against inverse magnetic field ($1/B$) between 0-9 T. The inset is the MR data obtained between 0-9 T at 2 K, which is collected in different sample from the above one. (b) The corresponding FFT of (a), where two peaks (two Fermi pockets) can be identified. (c) Landau fan diagrams of the two SdH modes. Extrapolated linear fitting gives intercepts, as enlarged near zero in the inset. The low-frequency $\alpha$ mode presents a nontrivial Berry phase. (d) The amplitude analysis, deducing the effective mass. (e) The oscillating MR at different temperatures, revealing the energy splitting.

**Figure 3. ARPES measurement, explicitly evidencing the existence of a 2D state.** (a) Fermi surface of ZrSiS single crystal. (b) GX cut measured along cut #1 indicated in the top panel of (a) with 46 eV (i) and 66 eV (ii) photons, respectively. (c) XM cut measured along cut #2 indicated in the top panel of (a) with 46 eV (i) and 66 eV (ii) photons, respectively. (d) The photon energy-$k_{//}$ intensity map along cut #1 direction. (e) The photon energy-$k_{//}$ intensity map along cut #2 direction.

**Figure 4. DFT calculations with both monolayer and bulk of ZrSiS.** (a) Schematic crystalline model of the ZrSiS monolayer used for the calculation. (b) The calculated bulk band structure without the spin-orbit coupling. (c) The calculated Fermi surface patterns for a monolayer ZrSiS (bottom) and the bulk solid (top). A small pocket can be seen around X for the monolayer (blue arrow), which is seen in the ARPES study. (d) The angle-dependent group velocity of the electrons along the direction of the magnetic field. (e) The calculated electronic structure of the monolayer.

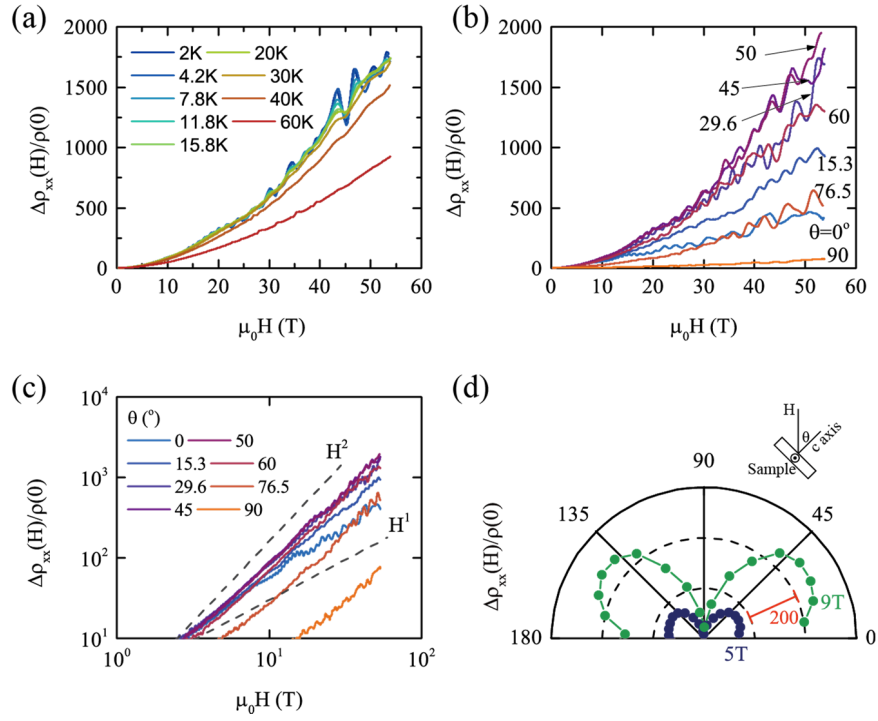

**Figure 1. Large, unsaturated and butterfly shaped anisotropic MR of the ZrSiS single crystals.**

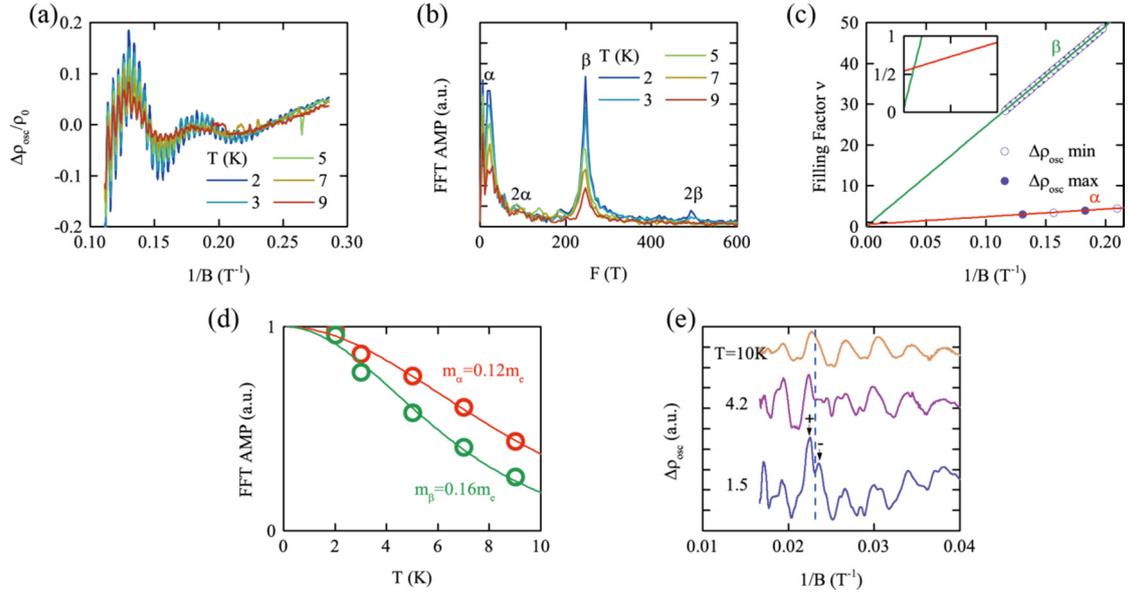

**Figure 2. Analyzing the SdH oscillations for a nontrivial Berry phase.**

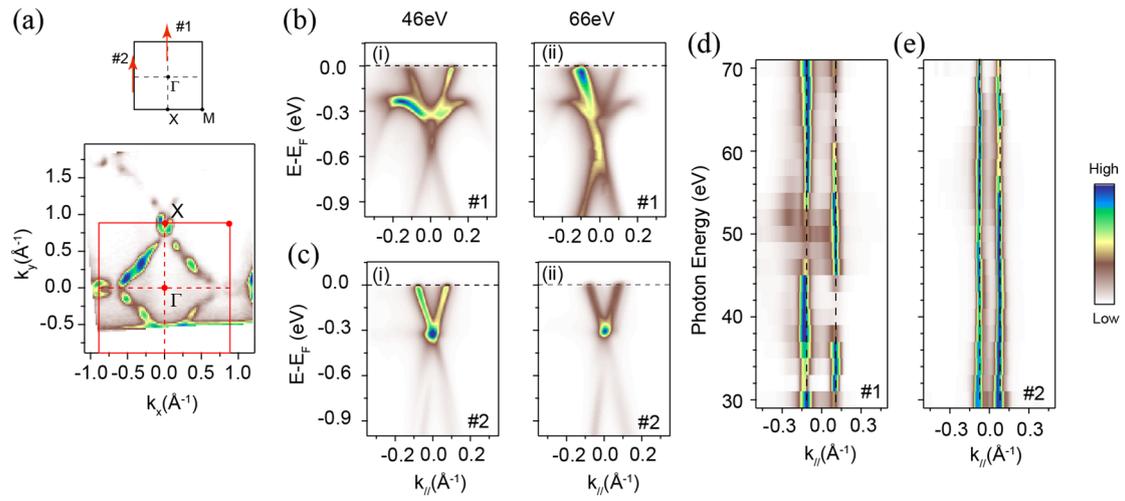

**Figure 3. ARPES measurement, explicitly evidencing the existence of a 2D state.**

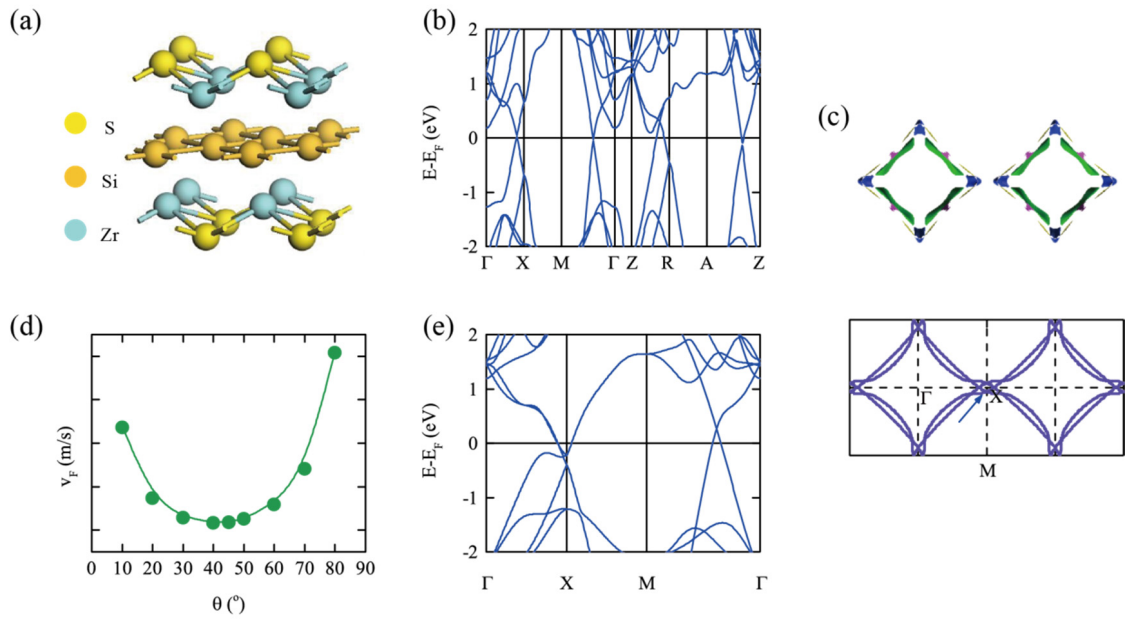

**Figure 4. DFT calculations with both monolayer and bulk of ZrSiS.**